\documentclass[12pt,journal,onecolumn,draftcls,peerreview]{IEEEtran}

\usepackage{amssymb}
\usepackage{amsthm}
\usepackage{algorithm}
\usepackage{algorithmic}
\usepackage{amsmath}
\allowdisplaybreaks[4]
\usepackage{amsfonts}
\usepackage{graphicx}
\usepackage{makeidx}
\usepackage{cite}
\usepackage{float}
\usepackage{epstopdf}
\usepackage{subfigure}
\usepackage{array}

\newtheorem{theorem}{\textbf{Theorem}}

\newtheorem{corollary}{\textbf{Corollary}}

\newtheorem{definition}{Definition}

\newtheorem{lemma}{\textbf{Lemma}}

\newtheorem{remark}{\textbf{Remark}}

\IEEEoverridecommandlockouts

\graphicspath{{figures/}}
\DeclareMathSizes{10}{10}{7}{3}

\begin{document}
%

\title{Optimal Multi-User Scheduling of Buffer-Aided Relay Systems}

\author{
\IEEEauthorblockN{Pihe Hu, Cheng Li, Dingjie Xu, Bin Xia }

\IEEEauthorblockA{Department of Electronic Engineering, Shanghai Jiao Tong University, Shanghai, China}

Emails: \{hupihe, lichengg, xudingjie1993, bxia\}@sjtu.edu.cn

\thanks{This work was supported in part by the the Huawei HIRP Project under Grant YB2015040062,  the National Nature Science Foundation of China under Grants 61771307, 61531009 and 61521062.}
}

\maketitle

\begin{abstract}
Multi-User scheduling is a challenging problem under the relaying scenarios. Traditional schemes, which are based on the instantaneous signal-to-interference-plus-noises ratios (SINRs), cannot solve the inherent disparities of the qualities between different links. Hence, the system performance is always limited by the weaker links. In this paper, from the whole system throughput view, we propose an optimal multi-user scheduling scheme for the multi-user full-duplex (FD) buffer aided relay systems. We first formulate the throughput maximization problem. Then, according to the characteristics of the Karush-Kuhn-Tucker conditions, we obtain the optimal decision functions and the optimal weighted factors of different links of the proposed scheme. Simulation results show that the proposed scheme not only solves the disparities of the qualities between $S_i$-$R$ and $R$-$D_i$ links, but also that between different $S_{i}$-$R$ or $D_{i}$-$R$ links, which can be used as guidance in the design of the practical systems.
\end{abstract}

\section{Introduction}
Relaying communication, which is an efficient way to improve the quality of service of wireless communication systems, has been investigated extensively. Practical field tests under the long term evolution (LTE) systems have shown that the relay nodes can eliminate the coverage holes of the macro base stations\cite{6239938}, as well as enhance the outdoor to indoor signal strengths \cite{6666597}. However, most of the existing works about the relaying communication only considered the single user scenarios \cite{7568989,7503846}, i.e., a source node communicate with a destination node via a relay. However, in the design of the fifth generation (5G) wireless communication systems, the volume of users is envisioned to improve 100 times compared to that of LTE systems \cite{4984446,7920321}. Hence, it is urgent to investigate the multi-user relay systems.

In the multi-user systems, the user scheduling schemes are crucial to the system performances. For this topic, in \cite{5599948}, the authors proposed the Max-link selection scheme and analyzed the average achievable rate with the consideration of the independent and identically distributed (i.i.d.) fading channels. To model the distance disparities between different users to the relay node, in \cite{7778253,7801128,8246574}, the authors have proposed two multi-user scheduling schemes and analyzed the system performances for the two-way full-duplex (FD) relay systems under the independent but not identically distributed (i.ni.d.) fading environment. However, the results revealed that the system performances were limited by the worse one of the $S$-$R$ and $R$-$D$ links of the selected users. In this case, the potentials of the good links could not be fully exploited. In addition, these works did not consider the buffer at the relay node\cite{7932158}, which led to the fixed transmission mode, i.e., the source transmission phase was always followed by the relay transmission phase.

Thereafter, the buffer was considered in the multi-user relay systems \cite{7555309, 7457724}. However, the multiple access channels were considered rather than the multi-user scheduling schemes. In \cite{7470977}, the authors have proposed a multi-user scheduling scheme for the buffer aided multi-user relay system, in which the relay node could achieve the adaptive link selection. In \cite{6177989}, the authors have investigated the system performance of three different multi-user scheduling schemes, i.e., Max-Min, Max-Max and Max link selection schemes, for the multi-relay systems. However, even equipped with buffers, the disparities of the qualities of different links have not been solved. As a consequence, the system performance was always restricted by the weaker links.

From the existing works about the multi-user relay systems, we observe that how to solve the disparities between different links remains unsolved. In this paper, we are dedicated to design an optimal multi-user scheduling scheme from the system throughput view for the FD multi-user relay systems. To solve this problem, we first formulate the throughput maximization problem as a binary integer optimization problem, which is known as an NP-hard problem. By relaxing the binary variables, we obtain the  Karush-Kuhn-Tucker (KKT) optimal conditions, based on which, we further obtain the optimal decision functions and optimal weighted factors of different links. To verify the proposed scheme, numerical simulations are conducted. The results indicate that the proposed scheme not only solve the disparities between the $S_{i}$-$R$ and $R$-$D_{i}$ links, but also that between different $S_{i}$-$R$($D_{i}$-$R$) links. In addition, the superiorities of the proposed scheme over the traditional schemes are revealed.

\section{System Model}
In this section, we elaborate on the physical-layer channel mode, specific CSI requirements and basic transmission scheme, respectively.

\begin{figure}[ht]
  \centering
  \includegraphics[width=4.0in]{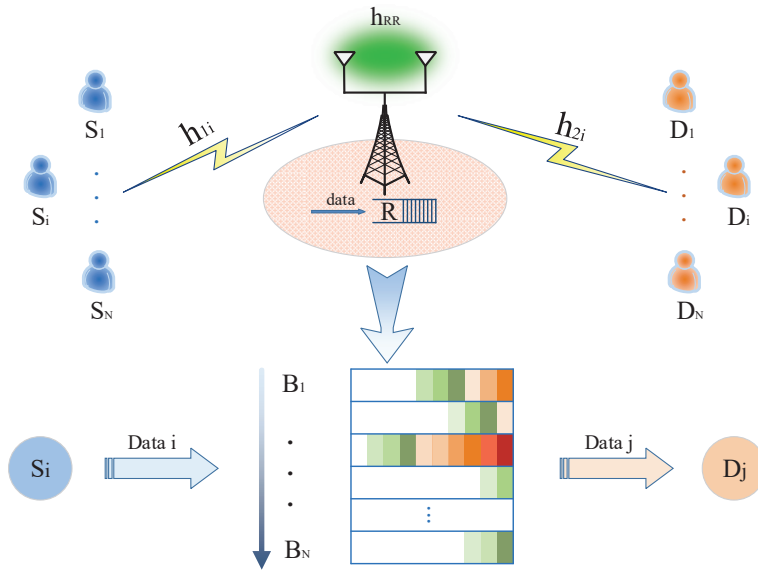}\\
  \caption{The multi-user buffer aided relay system, where the source $S_{i}$ sends messages to $D_{i}$ via the relay node.  The relay node is equipped with $N$ buffers from $B_{1}$ to $B_{N}$. The messages from the source $S_{i}$ are temporarily stored in the buffer $B_{i}$, where $i\in\{ 1, ... ,  N\}$.}
  \label{model}
\end{figure}

\vspace{-1mm}
\subsection{Channel Model}
As shown in the Fig.1, we consider a multi-user buffer aided FD relay system. The multiple users are pre-paired, i.e., $S_{i}$ exclusively sends messages to $D_{i}$. We assume that the direct link between $S_{i}$ and $D_{i}$ does not exist due to the strong fading caused by the large separation, penetration loss of buildings and strong shadowing effects. The relay node $R$ has the capability to work in the FD mode and is equipped with a large buffer, which is divided into $N$ sub-buffers from $B_{1}$ to $B_{N}$. The bit messages from the source $S_{i}$ are first stored in the buffer $B_{i}$, and then forwarded to the destination $D_{i}$ in the subsequential time slots.

We denote the channel coefficients of the $S_{i}$-$R$ link and $R$-$D_{i}$ link in the \emph{t-th} time slot as $h_{1i}(t)$ and $h_{2i}(t)$, respectively. We assume that $h_{1i}(t)$ and $h_{2i}(t)$ are subject to the i.ni.d., stationary and ergodic stochastic distributions. In addition, we consider the block fading channels, i.e., the channel coefficients keep constant within one time slot but change independently between different time slots. We denote the transmit powers of the sources and the relay as $P_{s}$ and $P_{r}$, respectively. Hence, the instantaneous signal-to-noise-ratios (SNRs) of the $S_{i}$-$R$ and $R$-$D_{i}$ links in the \emph{t-th} time slot are given by $s_{i}(t)\triangleq \gamma_{s}g_{1i}(t)$ and $r_{i}(t)\triangleq \gamma_{r}g_{2i}(t)$, with probability density functions (pdfs) $f_{s_{i}}(s)$ and $f_{r_{i}}(r)$, respectively. Here, $g_{1i}(t)=|h_{1i}(t)|^2$ and $g_{2i}(t)=|h_{2i}(t)|^2$ denote the channel gains of the $S_{i}$-$R$ and $R$-$D_{i}$ links, respectively. $\gamma_{s}=P_{s}/\sigma_{r}^2$ and $\gamma_{r}=P_{r}/\sigma_{d}^2$ denote the variances of the transmit SNRs of the source $S_{i}$ and the relay $R$, respectively. $\sigma_{r}^2$ and $\sigma_{d}^2$ denote variances the Gaussian noises at the relay node and the destination nodes, respectively. In addition, we denote the average SNRs as $\Omega_{s_{i}}\triangleq \mathbb{E}\{s_{i}(t)\}$ and $\Omega_{r_{i}}\triangleq \mathbb{E}\{r_{i}(t)\}$. At this point, we can give the capacity expressions of the $S_{i}$-$R$ and $R$-$D_{i}$ links as
\begin{align}\label{749464}
  C_{s_{i}}(t)&=\log_{2}(1+s_{i}(t)),\notag\\
  C_{r_{i}}(t)&=\log_{2}(1+r_{i}(t)).
\end{align}

\subsection{CSI Requirements}
To perform the optimal multi-user scheduling scheme, the knowledge of the global CSI is needed at the relay node. The global CSI can be obtained by the following steps. First, at the beginning of each time slot, all the source nodes transmit the pilot signals to the relay node $R$, and $R$ estimates $h_{1i}(t)$. Second, the relay node broadcasts the pilot signals to all the destinations, and each destination estimates $h_{2i}(t)$. Finally, all the destination nodes feed back the quantized version of $h_{2i}(t)$ to the relay node. In addition, according to the system parameters, including $P_{s}$, $P_{r}$, $\sigma_{r}^2$ and $\sigma_{d}^2$, the relay node can obtain the instantaneous channel capacities of all the links.

\subsection{Basic Transmission Scheme}
In the traditional multi-user relay systems with or without buffers, there are two scheduling schemes. First, for the system without buffer, in each time slot the relay node has to select one pair of users simultaneously. For instance, $S_{1}$ is selected to send messages and $D_{1}$ is selected to receive messages in the same time slot. Second, for the system with buffer, in each time slot, the relay node always chooses the source node with maximum $C_{s_{i}}(t)$ and the destination node with maximum $C_{r_{i}}(t)$. However, in this paper, we abandon these two schemes and propose an optimal user scheduling scheme. In the proposed scheme, the source node and the destination node are selected independently according to the selection criteria, which will be elaborated on in the Section III. Next, we briefly explain the basic transmission scheme.

If the source node $S_{i}$ is selected to send messages in the \emph{t-th} time slot, it transmits with the instantaneous rate
\begin{equation}\label{87343}
  R_{s_{i}}(t) = C_{s_{i}}(t).
\end{equation}
The messages will be first stored in the buffer $B_{i}$. We use $Q_{i}(t)$ to denote the buffer state, which will change to
\begin{equation}\label{876431}
  Q_{i}(t)=Q_{i}(t-1)+R_{s_{i}}(t).
\end{equation}

If the destination node $D_{i}$ is selected to receive message in the \emph{t-th} time slot, the relay node transmits with the instantaneous rate
\begin{equation}\label{213454}
  R_{r_{i}}(t)=\min\{Q_{i}(t-1), C_{r_{i}}(t)\}.
\end{equation}
The messages stored in the buffer $B_{i}$ will be sent to the destination $D_{i}$ and the buffer state will change to
\begin{equation}\label{872236}
  Q_{i}(t)=Q_{i}(t-1)-R_{r_{i}}(t).
\end{equation}

\section{Optimal User Scheduling Scheme}
In this section, we will first formulate the throughput maximization problem and then propose the optimal user scheduling scheme.
\subsection{Problem Formulation}
In this paper, we are dedicated to design an optimal user scheduling scheme for the buffer aided relay system under the i.ni.d. fading environment. Before that, we use the binary variables $p_{i}(t)$ and $q_{i}(t)$ to denote the indicator of which $S_{i}$ and $D_{i}$ are selected in the \emph{t-th} time slot. We have the following definition:

\begin{definition}
  For the source nodes
\begin{equation}
p_{i}(t)=\left\{
\begin{array}{ll}
\vspace{1mm}
1, &\text{the node $S_{i}$ is selected} \\
0, & \text{otherwise}
\end{array}
\right.
\end{equation}

For the destination nodes
\begin{equation}
q_{i}(t)=\left\{
\begin{array}{ll}
\vspace{1mm}
1, &\text{the node $D_{i}$ is selected} \\
0, & \text{otherwise}
\end{array}
\right.
\end{equation}
\end{definition}
It is clear that
\begin{align}\label{78764354}
  &\sum_{i=1}^{N} p_{i}(t) =1,\notag\\
  &\sum_{i=1}^{N} q_{i}(t)=1,
\end{align}
which guarantees that only one source node and one destination node are selected simultaneously in each time slot. In addition, we assume that all the source nodes have enough backlogged messages to transmit. Thus the average transmission rate of the source $S_{i}$ is given by
\begin{equation}\label{786413}
  R_{s_{i}}=\lim_{T\rightarrow \infty} \frac{1}{T}\sum_{t=0}^{T}p_{i}(t)R_{s_{i}}(t),
\end{equation}
and the maximum average reception rate at the destination node $D_{i}$ is given by
\begin{equation}\label{87643}
 R_{r_{i}} = \lim_{T\rightarrow \infty} \frac{1}{T}\sum_{t=0}^{T}q_{i}(t)R_{r_{i}}(t),
\end{equation}

The system throughput can be explained as the sum of the average reception rates of all the destination nodes. It is given by
\begin{align}\label{4876820}
  \mathcal{T}&=\sum_{i=1}^{N}R_{r_i}=\lim_{T\rightarrow \infty}\frac{1}{T}\sum_{i=1}^{N}\sum_{t=0}^{T} q_{i}(t)R_{r_{i}}(t).
\end{align}

Since we have considered a practical system, the buffer at the relay node has to satisfy the stationary conditions, which is given by
\begin{lemma}
 For a stationary buffer aided relay system, the average arrival rate and the departure rate have to satisfy
 \begin{equation}\label{1318545}
   R_{s_{i}}\leq R_{r_{i}}, \ \ i\in\{1,...,N\}
 \end{equation}
\end{lemma}
\begin{proof}
In order to be stationary, the length of the queue in the buffer cannot increase infinitely. However, according to the queueing theory, if $R_{s_{i}}> R_{r_{i}}$, the length of the queue will approach to infinity. Hence, we have Lemma 1.
\end{proof}

In addition to the stationary condition, in order to maximize the system throughput, we have the following lemma
\begin{lemma}
To maximize the throughput, the buffer at the relay node have to be at the edge of non-absorbing state, i.e.,
\begin{equation}\label{1537}
  R_{s_{i}} = R_{r_{i}}, \  \ i\in\{1,...,N\}
\end{equation}
\end{lemma}
\begin{proof}
  If the buffer is at the non-absorbing state, i.e., $R_{s_{i}} < R_{r_{i}}$, the system throughput is limited by $R_{s_{i}}$. Intuitively, we have two methods to improve the system throughput. One is that we can choose $S_{i}$ in more time slots to increase $R_{s_{i}}$. The other is that we can choose other node except $D_{i}$ in more time slots to decrease $R_{r_{i}}$. Once $R_{s_{i}}=R_{r_{i}}$, $R_{s_{i}}$ cannot be increased further more according to the Lemma 1.
\end{proof}

With this lemma, we can observe that the minimal limitation in (\ref{213454}) can be neglected. Because according to the queuing theory, when the average arrival rate equals to the average departure rate, the average queue length can be large enough. This means the relay node always have enough messages stored in the buffer, which can also be explained as that over the infinite time horizon, there are only accountable time slots that the buffer has insufficient messages to transmit. Hence, we can simplify the system throughput as
\begin{align}\label{167484641}
  \mathcal{T}=\lim_{T\rightarrow \infty}\frac{1}{T}\sum_{i=1}^{N}\sum_{t=0}^{T} q_{i}(t)C_{r_{i}}(t),
\end{align}
Now, we are ready to present the throughput maximization problem $\mathcal{P}_{1}$
\begin{align}\label{32481212}
  \mathop {\text{max}} \limits_{p_{i}(t), q_{i}(t)}&\qquad\quad \mathcal{T}  \\
  \text{s.t.}\quad &\text{C1 :} \quad R_{s_{i}}=R_{r_{i}},\quad \forall\ i\notag\\
  &\text{C2 :}\quad \sum_{i=1}^{N}p_{i}(t)=1,\quad\forall\ t\notag\\
  &\text{C3 :}\quad \sum_{i=1}^{N}q_{i}(t)=1, \quad\forall\ t\notag\\
  &\text{C4 :}\quad p_{i}(t)\in\{0, 1\}, \quad\forall\ i,\ t\notag\\
  &\text{C5 :}\quad  q_{i}(t)\in\{0, 1\}, \quad\forall\ i,\ t
\end{align}
where C1 ensures that every sub-buffer is at the edge of non-absorbing state. C2 and C3 guarantee that only one source and one destination node are selected during one time slot. C4 and C5 ensure that there are only two states of a source node and a destination node, i.e., selected or not selected.

We note that the original problem $\mathcal{P}_{1}$ is a binary integer programming problem, which is NP-hard. To solve this problem, we relax the binary decision variables $p_{i}(t)$ and $q_{i}(t)$ to be continuous ones varying between 0 and 1, i.e., $p_{i}(t)$, $q_{i}(t) \in [0,1]$. The problem after relaxation is denoted by $\mathcal{P}_{2}$,

\begin{align}\label{32481}
  \mathop {\text{max}} \limits_{p_{i}(t), q_{i}(t)}&\qquad\quad \mathcal{T}  \\
  \text{s.t.}\quad &\text{C1 :} \quad R_{s_{i}}=R_{r_{i}},\quad \forall i\notag\\
  &\text{C2 :}\quad \sum_{i=1}^{N}p_{i}(t)=1,\quad\forall t\notag\\
  &\text{C3 :}\quad \sum_{i=1}^{N}q_{i}(t)=1, \quad\forall t\notag\\
  &\text{C4 :}\quad 1-p_{i}(t) \ge 0, \quad\forall i,\ t\notag\\
  &\text{C5 :}\quad p_{i}(t) \ \ge \ 0, \qquad\ \ \forall i,\ t \notag\\
  &\text{C6 :}\quad 1-q_{i}(t) \ge 0, \quad \forall i,\ t \notag\\
  &\text{C7 :}\quad  q_{i}(t) \ \ge\ 0, \qquad\ \ \forall i,\ t
\end{align}

\subsection{Optimal Multi-User Scheduling Scheme}
Here we present the optimal multi-user scheduling scheme.
\begin{theorem}
  The optimal multi-user scheduling scheme maximizing the system throughput is given by

  For the source nodes:
\begin{equation}\label{74684}
    p_{k}(t)=\left\{\begin{array}{l}
               1,\quad k =\text{ arg} \max\limits_{i=1,\cdots, N} \Gamma_{i}(t)\\
               0,\quad \text{otherwise}
             \end{array}
             \right.
  \end{equation}

  For the destination nodes:
  \begin{equation}\label{74684111}
    q_{k}(t)=\left\{\begin{array}{l}
               1,\quad k =\text{ arg} \max\limits_{i=1,\cdots, N} \Lambda_{i}(t)\\
               0,\quad \text{otherwise}
             \end{array}
             \right.
  \end{equation}
  where the optimal decision functions are given by
  \begin{align}\label{4648422}
    \Gamma_{i}(t) &=-\lambda_{i}\mathcal{F}(s_{i}(t))\\
    \Lambda_{i}(t)&=(1+\lambda_{i})\mathcal{F}(r_{i}(t))
  \end{align}
  in which
  \begin{equation}\label{78644543}
    \mathcal{F}(x)=\log_{2}(1+x)
  \end{equation}
  denotes the optimal metric function and $\lambda_{i}\in(-1,0)$ denotes the weighted factors of the link $S_{i}-R$, which satisfy
  \begin{equation}\label{13657}
  \mathbb{E}\{p_{i}(t)C_{s_{i}}(t)\}=\mathbb{E}\{q_{i}(t)C_{r_{i}}(t)\}
  \end{equation}
\end{theorem}

\begin{proof}
  Please see Appendix A.
\end{proof}
\begin{remark}
We note that after relaxation, the original problem converts to a $2N$-dimensional linear programming problem over an infinite time horizon. For this problem, the optimal solution always locates at the boundary of the feasible set, i.e., $p_{i}(t), q_{i}(t)$ = 0 or 1, which coincides with the original problem. Hence, the optimal solution of the problem $\mathcal{P}_{2}$ is also the optimal solution of the original problem.
\end{remark}
\begin{remark}
  It is noted that this scheme is particularly suitable for the i.ni.d. fading environment. Unlike the traditional multi-user scheduling scheme, in which the system performance is limited by the worse one of the $S$-$R$ and $R$-$D$ links, the proposed scheme can balance the quality of the $S_i$-$R$ and $R$-$D_i$ links by introducing the weighted factor $\lambda_{i}$. For instance, we assume that the expectation of the channel gain of the first pair are $\Omega_{s_1}=10$ and $\Omega_{r_1}=1$. We may set $\lambda_{1}=-0.1$ and $1+\lambda_{1}=0.9$, which is equivalent to that the $S$-$R$ link is weakened with factor $0.1$ but the $R$-$D$ link is weakened with factor $0.9$. The source node $S_{1}$ will be selected in less time slots and the destination node $D_{1}$ will be selected in more time slots. By this way, the average arrival rate and the average departure rate of the buffer $B_{1}$ can be balanced.
\end{remark}


\section{Maximum System Throughput}

To obtain the maximum system throughput, we need to first obtain the $\lambda_{i}$, which is given by the following corollary
\begin{corollary}
For the proposed optimal scheduling scheme in theorem 1, the optimal weighted factors are the solutions of the following equations
\begin{align}\label{57027548}
  &\int_{0}^{\infty}\Big[\prod_{j=1, j\neq i}^N\int_{0}^{\mathcal{H}_{s_{j}}^i}f_{s_{j}}(s_{j})dt\Big]\log_{2}(1+s_{i})f_{s_{i}}(s_{i})ds_{i} \notag\\
  =&\int_{0}^{\infty}\Big[\prod_{j=1, j\neq i}^N\int_{0}^{\mathcal{H}_{r_{j}}^i}f_{r_{j}}(r_{j})dr_{j}\Big]\log_{2}(1+r_{i})f_{r_{i}}(r_{i})dr_{i}
\end{align}
where
\begin{align}\label{4768746341}
  \mathcal{H}_{s_{j}}^i & =\mathcal{F}^{-1}(\frac{\lambda_{i}}{\lambda_{j}}\mathcal{F}(s_{i})) \\
  \mathcal{H}_{r_{j}}^i & =\mathcal{F}^{-1}(\frac{1+\lambda_{i}}{1+\lambda_{j}}\mathcal{F}(r_{i}))
\end{align}
\end{corollary}
\begin{proof}
  Please see Appendix B.
\end{proof}

We can obtain the optimal weighted factor $\lambda_{i}$ using the built-in functions of the software packages such as Matlab or Mathmetica. Then the system throughput is given by the following Theorem.

\begin{theorem}
Based on the theorem 1 and the corollary 1, the system throughput can be easily obtained as
\begin{equation}\label{4986741}
  \mathcal{T}=\sum_{i=1}^{N}\int_{0}^{\infty}\!\Big[\!\!\prod_{j=1, j\neq i}^N\!\!\int_{0}^{\mathcal{H}_{r_{j}}^i}\!\!\!f_{r_{j}}(r_{j})dr_{j}\Big]\log_{2}(1+r_{i})f_{r_{i}}(r_{i})dr_{i}
\end{equation}
\end{theorem}
\begin{proof}
  The system throughput can be easily obtained using the Total Probability Theorem, the details of the proof are omitted here.
\end{proof}

\section{Simulation Results}
In this section, we present the simulation results to verify the proposed multi-user scheduling scheme. For simplicity and without loss of generality, we consider two pairs of users in the system. In the simulation, we adopt the Rayleigh fading environment with SNR expectations $E\{s_{i}(t)\}=\Omega_{s_{i}}$, $E\{r_{i}(t)\}=\Omega_{r_{i}}$, $i \in \{1,2\}$. In addition, we also present the sub-optimal method, in which we use $\mathcal{F'}(x)=x$ to approximate $\mathcal{F}(x)=\log_{2}(1+x)$ to reduce the complexity in the simulation. As a results, the sub-optimal weighted factors will be obtained.

In the Fig. 2, we plot the optimal and sub-optimal weighted factors  $\lambda_{i}$ versus the channel variation. We set $\Omega_{s_{2}}=4$, $\Omega_{r_{2}}=6$ and $\Omega_{s_{1}}+\Omega_{r_{1}}=10$, and change the ratio of $\Omega_{s_{1}}/\Omega_{r_{1}}$. The simulation results verify that the value of $\lambda_{i}$ varies between -1 and 0. In addition, we note that with the increase of the ratio $\Omega_{s_{1}}/\Omega_{r_{1}}$, $\lambda_{1}$ increases from -1 to 0. Actually, this is consistent with our intuition, the larger $\lambda_{1}$ represents that the $S_{1}$ will be selected in less time slots. Furthermore, the variation of $\Omega_{s_{1}}/\Omega_{r_{1}}$ will also affects the value of $\lambda_{2}$. When $\Omega_{s_{1}}/\Omega_{r_{1}}$ is very large, $\lambda_{2}$ will approach to -1, which implies that when $S_{1}$ is selected in less time slots, then $S_{2}$ will be selected in more time slots. This insight reveals the advantages of our proposed multi-user scheduling scheme. By multiplying different weighted factors, we can adjust the selection probabilities of different nodes, and by this way, to balance the disparities of the channel qualities between different $S_{i}$-$R$ links.

\begin{figure}[ht]
  \centering
  \includegraphics[width=3.5 in]{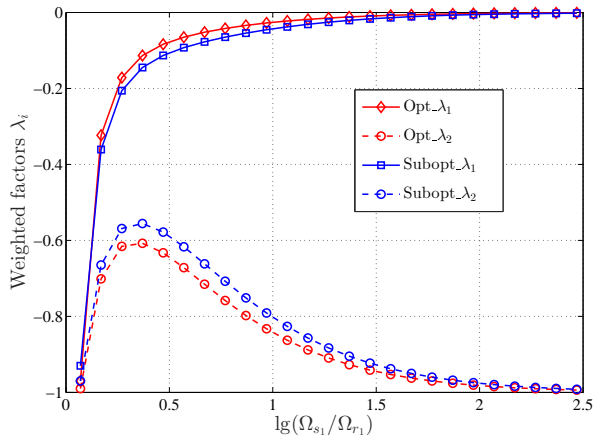}
  \vspace{0mm}
  \caption{System throughput vs. channel variation.}\label{4796851}
\end{figure}

\begin{figure}[ht]
  \centering
  \includegraphics[width=3.5 in]{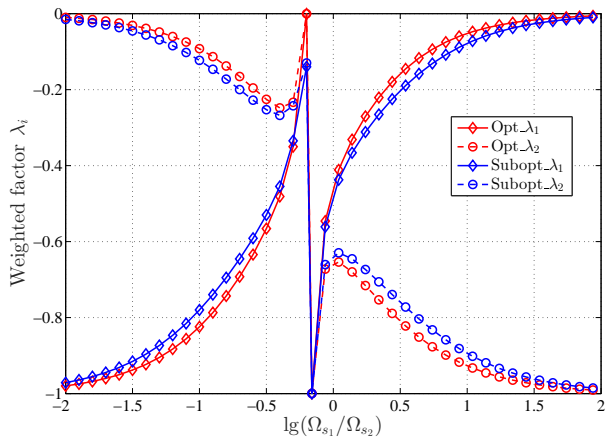}
  \vspace{0mm}
  \caption{system throughput $\mathcal{T}$ vs. channel variation.}\label{47968453211}
\end{figure}

Fig. 3 plots the optimal and sub-optimal weighted factor $\lambda_{i}$ with the variation of the ratio $\Omega_{s_{1}}/\Omega_{s_{2}}$. We set $\Omega_{r_{1}}=4$, $\Omega_{r_{2}}=6$ and $\Omega_{s_{1}}+\Omega_{s_{2}}=10$. This figure shows the relationship between the variation of the weighted factor $\lambda_{1}$ and $\lambda_{2}$. The results reveals that in the low $\Omega_{s_{1}}/\Omega_{s_{2}}$ regime, i.e., the $S_{1}$-$R$ link is weaker than the $S_{2}$-$R$ link, $\lambda_{1}$ is close to -1, but $\lambda_{2}$ is close to 0. This will lead to that $S_{1}$ will be selected in more time slots than $S_{2}$. This insights reveals another advantage of the proposed scheme, i.e., the ability to balance the disparities of the channel qualities between $S_{i}$-$R$ and $S_{j}$-$R$, $i\neq j$. In addition, we note that when $\Omega_{s_{1}}/\Omega_{s_{2}}\approx 2/3$, i.e., $\Omega_{s_{1}}=4$ and $\Omega_{s_{2}}=6$, $\lambda_{1}$ and $\lambda_{2}$ almost have the same value and the curves are almost completely vertical, which means any value of $\lambda_{1}=\lambda_{2}$ is optimal. In this case, $S_{i}$-$R$ and $R$-$D_{i}$ links have the same quality. The proposed scheme will convert to the traditional Max-Max user scheduling scheme.

\begin{figure}[ht]
  \centering
  \includegraphics[width=3.5in]{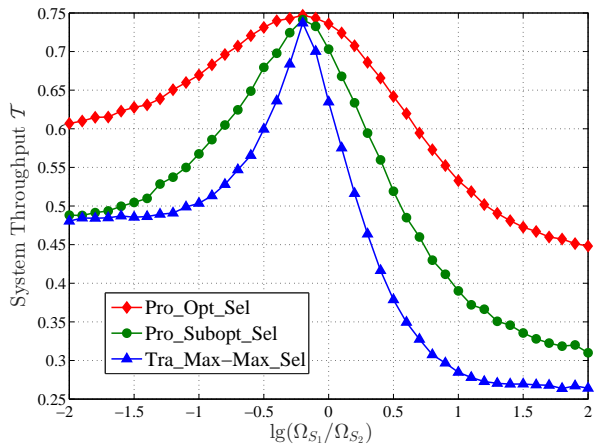}
  \vspace{0mm}
  \caption{system throughput $\mathcal{T}$ vs. channel variation.}\label{479684565431}
\end{figure}

Fig. 4 plots the system throughput versus the variation of the ratio $\Omega_{s_{1}}/\Omega_{s_{2}}$. The figure shows that the proposed user scheduling scheme achieves considerable performance gain over the traditional Max-Max user scheduling scheme. The reason is that under the traditional Max-Max scheme, the system has no ability to balance the disparities of the qualities between different links. Hence, the throughput of each user pair is limited by the weaker one of the $S_{i}$-$R$ and $R$-$D_{i}$ links. However, the proposed scheme can schedule more time slots to the weaker links and less time slots to the stronger links. In this way, the system throughput can be improved. In addition, we note that when $\Omega_{s_{1}}/\Omega_{s_{2}}\approx 2/3$, i.e., $\Omega_{s_{1}}=4$ and $\Omega_{s_{2}}=6$, the three different schemes have the same system throughput. This is consistent with the results in the Fig. 3. In this case, the proposed scheme converts to the traditional Max-Max scheme, which leads to the same performance.


\section{Conclusion}
In this paper, we proposed an optimal multi-user scheduling scheme for the FD multi-user buffer aided relay system. The proposed scheme could solve the disparities of the qualities between different links and maximize the system throughput. To obtain this scheme, we formulated the throughput maximization problem as an binary integer optimization problem. By relaxing the binary variables and based on the KKT optimal conditions, we obtained the optimal decision functions and optimal weighted factors of different links. Simulations were conducted and the results verified the effectiveness and advantages of the proposed scheme over the traditional schemes.

\appendices
\section{Proof of Theorem 1}
We note that the problem $\mathcal{P}_{2}$ is a 2N-dimensional linear programming problem over infinite time horizon. For this problem, the optimal solution satisfies the KKT optimal conditions. Hence, the lagrangian function can be expressed as
\begin{eqnarray}
   & & \mathcal{L}(p_{i}(t),q_{i}(t),\lambda_{i},\alpha(t),\beta(t),\mu_{i}(t),\nu_{i}(t),\zeta_{i}(t),\eta_{i}(t)) \notag \\
  &=&\frac{1}{T}\sum_{i=1}^{N}\sum_{t=0}^{T}q_{i}(t)C_{r_{i}}(t)\notag\\
  &-&\frac{1}{N}\sum_{t=0}^{T}\lambda_{i}\big[p_{i}(t)C_{s_{i}}(t)-q_{i}(t)C_{r_{i}(t)}\big]\notag\\
  &-& \alpha(t)\big[1-\sum_{i=1}^{N}p_{i}(t)\big]-\beta(t)\big[1-\sum_{i=1}^{N}q_{i}(t)\big]\notag\\
  &+&\mu_{i}(t)[1-p_{i}(t)]+\nu_{i}(t)p_{i}(t)\notag\\
  &+&\zeta_{i}(t)[1-q_{i}(t)]+\eta_{i}(t)q_{i}(t)
\end{eqnarray}
where $\lambda_{i}$, $\alpha(t)$, $\beta(t)$, $\mu_{i}(t)$, $\nu_{i}(t)$, $\zeta_{i}(t)$ and $\eta_{i}(t)$ are the Lagrange multipliers. Differentiate the $\mathcal{L}(\cdot)$ function with regard to $p_{i}(t)$ and $q_{i}(t)$, and set them to zero, i.e.,
\begin{equation}\label{748788}
  \frac{\partial \mathcal{L}}{\partial p_{i}(t)} = 0,\ \frac{\partial \mathcal{L}}{\partial q_{i}(t)} =0,
\end{equation}
we get
\begin{align}\label{1338413}
   & -\lambda_{i}\frac{1}{N}C_{s_{i}}(t)+\alpha(t)-\mu_{i}(t)+\nu_{i}(t)=0, \notag \\
   & (1+\lambda_{i})\frac{1}{N}C_{r_{i}}(t)+\beta(t)-\zeta_{i}(t)+\eta_{i}(t)=0,
\end{align}
respectively.

If we let $p_{k}(t)=1$, we obtain that $\nu_{k}(t)=0$, $\mu_{i}(t)=0$, $i \neq k$, according to the complementary slackness theorem. Substituting them into (\ref{1338413}), we have
\begin{eqnarray}\label{48764513541}
  N[-\alpha(t)+\mu_{k}(t)]& =& -\lambda_{k}C_{s_{k}}(t) \triangleq \Gamma_{k}(t)\nonumber\\
  N[-\alpha(t)-\nu_{i}(t)] &=& -\lambda_{i}C_{s_{i}}(t)\triangleq \Gamma_{i}(t),\ \  i\neq k
\end{eqnarray}

Since $\mu_{k}(t)\ge0$, $\nu_{i}(t)\ge0$, we have $\Gamma_{k}(t)\ge\Gamma_{i}(t)$, $i\neq k$, where $\Gamma_{i}(t)$ denotes the optimal decision function for the source nodes. Here, we can conclude that if $p_{k}(t)=1$, $\Gamma_{k}(t)\ge\max \{\Gamma_{i}(t)\}$. Hitherto, we obtain the optimal scheduling scheme for the source nodes. In the similar way, if we set $q_{k}(t)=1$, we have
\begin{eqnarray}
  N[-\beta(t)+\zeta_{k}(t)] &=& (1+\lambda_{k})C_{r_{k}}(t)=\Lambda_{k} \nonumber\\
  N[-\beta(t)-\eta_{i}(t)] &=& (1+\lambda_{i})C_{r_{i}}(t)=\Lambda_{i}
\end{eqnarray}
where $\zeta_{k}(t)\ge 0$, $\eta_{i}(t)\ge 0$, we have $\Lambda_{k}(t)\ge\Lambda_{i}(t), i\neq k$, which denotes the optimal decision function for the destination nodes. Hitherto, we have the necessary conditions of the  optimal scheduling scheme for the destination nodes.

Next, we need to specify the range of the optimal weighted factors $\lambda_{i}$. First, in order to make the buffers at the edge of non-absorb state, the average arrival rate should equal to the average departure rate, we have
\begin{equation}\label{16468441}
\mathbb{E}\{p_{i}(t)C_{s_{i}}(t)\}=\mathbb{E}\{q_{i}(t)C_{r_{i}}(t)\}.
\end{equation}

Second, if $\lambda_{i}= 0$ and $(1+\lambda_{i})= 0$, the corresponding source node $S_{i}$ and $D_{i}$ will never be selected. Hence, we have $\lambda_{i}\notin \{-1,0\}$. Finally, we consider a special case, i.e., i.i.d. fading environment, which means all the links have the same link quality, i.e., completely symmetric. In this case, $\lambda_{1}=\lambda_{2}=\cdots =\lambda_{N}=c_{0}$, where $c_{0}$ is a constant. In order to maximize the system throughput, the source with the maximum $C_{s_{i}}(t)$, say $S_{k_1}$ and the destination with maximum $C_{r_{j}}(t)$, say $D_{k_2}$ will be selected. Thus, we have that $-\lambda_{k_1}>0$ and $-(1+\lambda_{k_2})>0$. Hence, we can conclude that $\lambda_{i}\in(-1,0)$.

\section{Proof of Corollary 1}
First, we define $2N$-dimensional random variables, $\mathbf{s(t)}=\{s_{1}(t), s_{2}(t),\cdots, s_{N}(t)\}$ and $\mathbf{r(t)}=\{r_{1}(t), r_{2}(t),\cdots, r_{N}(t)\}$, of which the \emph{pdfs} are
\begin{equation}\label{78767243}
  f_{\mathbf{s}}(\mathbf{s})=\prod_{i=1}^{N} f_{s_{i}}(s_{i}),
\end{equation}
\begin{equation}\label{7873467243}
  f_{\mathbf{r}}(\mathbf{r})=\prod_{i=1}^{N} f_{r_{i}}(r_{i}).
\end{equation}

Hence, for $i=1,2,\cdots,N$, we have
\begin{align}\label{79843415}
\mathbb{E}\{p_{i}(t)C_{s_{i}}(t)\}
&=\int \limits_{-\lambda_{j}\mathcal{F}(s_{j})<-\lambda_{i}\mathcal{F}(s_{i})} \!\!\!\!\!\!f_{\mathbf{s}}(\mathbf{s})\log_{2}(1+s_{i})d\mathbf{s}\notag\\
&=\int_{0}^{\infty}\Big\{\Big[\prod_{j=1,j\neq i}^{N}\int_{0}^{\mathcal{H}_{s_{j}}^i}f_{s_{j}}(s_{j})ds_{j}\Big]\notag\\
& \log_{2}(1+s_{i})f_{s_{i}}(s_{i})\Big\}
\end{align}
\begin{align}\label{57430921875}
\mathbb{E}\{q_{i}(t)C_{r_{i}}(t)\} & = \int \limits_{(1+\lambda_{j})\mathcal{F}(r_{j})<(1+\lambda_{i})\mathcal{F}(r_{i})} \!\!\!\!\!\!f_{\mathbf{r}}(\mathbf{r})\log_{2}(1+r_{i})d\mathbf{r}\notag\\
&=\int_{0}^{\infty}\Big\{\Big[\prod_{j=1,j\neq i}^{N}\int_{0}^{\mathcal{H}_{r_{j}}^i}f_{r_{j}}(r_{j})dr_{j}\Big]\notag\\
& \log_{2}(1+r_{i})f_{r_{i}}(r_{i})\Big\}
\end{align}
where $\mathcal{H}_{s_{j}}^i$ and $\mathcal{H}_{r_{j}}^i$ are given by (\ref{57027548}) and (\ref{4768746341}). Substituting (\ref{79843415}) and (\ref{57430921875}) into (\ref{13657}), we can get the corollary 1, which ends the proof.
\bibliographystyle{IEEEtran}
\bibliography{reference}
\end{document}